On Real and Virtual Photons in the Davies Theory

of Time-Symmetric Quantum Electrodynamics

R. E. Kastner

University of Maryland, College Park

October 20, 2016*

ABSTRACT. This paper explores the distinction between virtual and real photons in the context of the Davies quantum relativistic extension of the Wheeler-Feynman classical electromagnetic theory. An alternative way of understanding this distinction is proposed, based on the transactional picture as first proposed by Cramer.

1. Introduction

This paper is devoted to exploring the subtle issue of the distinction between virtual photons and real photons, with specific reference to the Davies quantum electrodynamics extension of the Wheeler-Feynman classical electromagnetic theory ((Davies 1970, 1971, 1972; Wheeler and Feynman 1945, 1949). I first briefly review the W-F theory, its application to the Transactional Interpretation of John G. Cramer (1986), and Davies' extension of the WF theory. I then suggest a slightly different approach from Davies in understanding the distinction between real and virtual photons in light of my recent application of his theory to the extension of the Transactional Interpretation into the relativistic domain. Finally, I point out the possibility that this alternative understanding of the real/virtual photon distinction might have empirically detectable consequences and could thus serve as a test of PTI.

2. The Wheeler-Feynman theory

The Wheeler-Feynman ('WF') theory proposes that radiation is fundamentally a time-symmetric process: a charge emits a field in the form of half-retarded, half-advanced solutions to the wave equation, and the response of absorbers combines with that primary field to create a radiative process that transfers energy from an emitter to an absorber. Specifically, when a given source emits its time-symmetric field, other charges respond to that field by emitting their own time-symmetric field, but exactly out of phase with the stimulating field. Using the additivity of radiation fields, Wheeler and Feynman show that, if the universe is a 'light-tight box'[1], the overall advanced response ('echo') of all absorbers to the retarded radiation from any particular emitter amounts to precisely the same field as that original half-strength retarded radiation field from the emitter.

---

*This is an updated version of the published paper (EJTP 11, #30, 2014), with a few corrections and clarifications.

The above process results in two distinct effects: (1) The two fields add; thus the retarded field from the source attains full strength and the advanced components cancel. (2) In addition, the absorber response provides for a 'free field' component that must be assumed in an *ad hoc* manner in the standard theory (which assumes that the source emits only a retarded field ) in order to account for the loss of energy by a radiating charge.[2] This two-fold process, wherein the advanced field from the absorber (1) superimposes constructively with the retarded field of an emitter and (2) provides for energy transfer from the emitter to the absorber, forms the basis for the 'transaction' in TI.

3. The Davies Theory

Davies (1970, 1971, 1972) presented an extension of the Wheeler-Feynman time-symmetric theory of electromagnetism to the quantum relativistic domain. This theory followed the basic Wheeler-Feynman method by introducing a boundary condition of a perfectly absorbing future universe in order to argue that the field due to a particular emitting point charge could be seen as composed of equal parts retarded radiation from the charge and advanced radiation from the absorbing systems.

The Davies papers demonstrate that the field due to a particular emitting current $j_{(i)}{}^\mu(x)$ can be seen as composed of equal parts retarded radiation from the emitting current and advanced radiation from absorbers. Using an S-matrix formulation, Davies replaces the action operator of standard QED,

$$J = \sum_i \int dx\, j^\mu{}_{(i)}(x) A_\mu(x) \qquad (1)$$

---

[1] This means that any emitted radiation is fully absorbed; no retarded radiation escapes to future infinity.

[2] The 'free field' is the difference of the retarded and advanced solutions. It has the properties of a field that does not arise from (or converge onto) a source (or sink), but simply exists independently. As such it is a solution to the homogeneous equation for the field.

(where $A_\mu$ is the standard quantized electromagnetic field), with an action derived from a direct current-to-current interaction,[3]

$$J = -\frac{1}{2} \sum_{i,j} \int dx\, dy\, j^\mu{}_{(i)}(x) D_F(x-y) j_{(j)\mu}(y), \qquad (2)$$

where $D_F(x-y)$ is the Feynman photon propagator. (This general expression includes both distinguishable and indistinguishable currents.)

While $D_F(x-y)$ imposes explicit temporal asymmetry in that it only allows positive frequencies to propagate into the future, Davies shows that for a 'light-tight box' (i.e., no free fields), the Feynman propagator can be replaced by the time-symmetric propagator $\overline{D}(x) = \frac{1}{2}\left[D^{ret}(x) + D^{adv}(x)\right]$, where the terms in the sum are the retarded and advanced Green's functions (solutions to the inhomogeneous wave equation); therefore the Davies theory and the standard theory are equivalent, at the level of the basic field propagation,[4] under these boundary conditions. This shows that the 'arrow of time' can arise from specific boundary conditions rather than needing to be imposed on the basic propagation of the fields themselves.

---

[3] That these expressions are equivalent is proved in Davies (1971) and reviewed in (1972).

[4] The Dirac theory of 1970-72 is formally equivalent to standard QED when absorption is assumed to occur universally. However, it readily allows for a transactional interpretation in which it is only for 'real' photons that there is an absorber response; this is what is explored herein.

The work by Davies provides a natural theoretical basis for the extension of the Transactional Interpretation (TI) of Cramer (1980, 1986) into the relativistic domain; this was presented in Kastner (2012a, b) in terms a 'possibilist' ontology (Possibilist Transactional Interpretation, 'PTI'). While the electromagnetic field is not quantized in Davies' direct-action theory, the latter can readily be studied in the 'Feynman diagram' technique provided that the virtual (internal) lines are understood as described by the time-symmetric propagator rather than by the Feynman propagator, and the 'external lines' are understood as 'real particles' which have been emitted with certainty and have received an absorber response (the external emitters/absorbers not being included in the diagram). In contrast, an internal line represents a virtual particle with an amplitude less than unity (i.e., the coupling constant) for both emission and absorption.[5]

4. Virtual vs. real photons

I begin this section by quoting from Davies (1972), for his discussion raises a subtle issue that is argued here needs a slightly different solution than the one he proposes.

"When we quantize the free electromagnetic field, we build up a Fock space out of states containing all numbers of photons. These photons obey the relation $k^2 = 0$. and, by the uncertainty principle, have an infinite lifetime. When the field is coupled to its sources, we allow for photons to be created and annihilated. If a photon is created at $t = 0$, and destroyed at $t = T$, we expect that it will lie off the energy shell (ie that $k^2 \neq 0$) for finite $T$. We say that such photons are virtual. However, this simple picture can be very misleading and confusing…" (Davies 1972, p.1027)

Davies then recalls his proof (Davies 1971) that his theory using the time-symmetric propagator and assuming full ultimate absorption (light-tight box) conditions is equivalent to standard QED

---

[5] It has been noted by an anonymous critic that the Davies theory is not Cauchy-complete. That is, there is no Cauchy complete inner product state space, i.e., Hilbert or Fock Space, for the basic propagating field. This is just another way of saying that the field is not quantized. But this is not a problem, especially in view of the possibility explored herein—that the virtual quanta (propagators) are subject to elevation into entities describable by a complete (Hilbert or Fock) space. Furthermore, at the meta-level, there is no compelling reason to assume that the physical universe has a Cauchy-complete structure, and to require on that basis that any field theory must be Cauchy-complete, especially in view of Gödel's Incompleteness Theorem.

with the Feynman propagator, as expressed in the following equivalence for the Scattering matrix,

$$<0|P \exp(-i \int J(x)A(x)\, dx\, |0> = P \exp\left(\tfrac{1}{2} i \int\int J(x)\, D_F(x-y)\, J(y)\, dx\, dy\right) \quad (3)$$

(with *P* the time-ordering operator). He then notes:

"By taking the [photon vacuum expectation value on the left-hand side of (3)] we indicate that the system has no real photons at $t = +/- \infty$. However, let us examine the photon propagator $D_F$ in detail. A Fourier decomposition gives

$$D_F(x) = \frac{1}{(2\pi)^4} \int \left( \frac{PP}{k^2} - i\pi\delta(k^2) \right) e^{ikx}\, dk = \overline{D} + D_1 \quad (4)$$

where PP is the principal part. The $\overline{D}$ part (bound field) leads to the real principal part term which describes virtual photons ($k^2 \neq 0$), whilst the imaginary $D_1$ (free field) describes photons with $k^2 = 0$, that is, real photons, through the $\delta$ function term. .. But how do we reconcile the notion of a real photon as an internal line in the Feynman diagram with the uncertainty principle? In other words, how can a real photon, which ought to have an infinite lifetime, be emitted and reabsorbed….? (*ibid.*)

Davies goes on to argue that both virtual and real photons can carry real energy, and that all 'real' photons have existed since $t = -\infty$. However, while the basic formulation of Davies' proposal is sound, this part of his argument is less convincing. The present author would like to suggest that it is not realistic to restrict the definition of a 'real' photon only to photons with an infinite lifetime. A more natural distinction between real and virtual photons is that a real photon is one that transfers empirically detectable energy, while virtual photons do not. Indeed the situation has a much more natural explanation in the transactional picture. In that picture, the response of the absorber is what gives rise to the 'free field' that in the quantum domain is considered a 'real photon'. So the 'realness' of the photon is defined in the transactional picture not by an infinite lifetime -- which, in reality, is practically never obtained – but rather by the presence of an absorber response. This response is what can give rise (through an actualized transaction) to a real photon with the ability to convey empirically detectable energy from one place to another, which is the function of the free field. That is, the idea that a 'real' photon must always have exactly zero rest mass is an idealization.

Thus, according to PTI, 'virtual' photons are those that do not convey empirically detectable energy between their source currents. In the Davies theory, these are described by the time-symmetric propagator $\bar{D}$, indicating that for virtual photons there is no absorber response. These virtual connections between currents (vertices in Feynman diagrams) are characterized by a coupling constant; in QED, it is $\sqrt{\alpha} \approx e$, the unit electric charge (in natural units). As observed in Kastner (2012a,b), the coupling constant is the amplitude for a charged particle to emit a photon. When generalized to the transactional picture, the coupling constant characterizes both the amplitude to emit an 'offer wave' (the usual quantum state) and the amplitude to generate a 'confirmation wave' (the advanced/dual quantum state). So in a typical Feynman diagram with a photon propagator (see Figure 1), both vertices express the *possibility* of emission of an offer and confirmation wave. *The virtual photon, in the transactional picture, is just this nascent possibility of a transaction between two currents—but one that was not realized*. A transaction is only attainable for virtual photons that satisfy the energy and momentum conservation constraints for the initial and final states of the system. Quantitatively, the probability of a transaction is $\alpha$ (the square of the coupling constant), times the square of the relevant matrix element. If such a transaction occurs, the virtual photon is elevated to a real photon, since it transfers real, positive energy from one object to another.

This solves the puzzle referred to by Davies, wherein the Feynman propagator -- characterizing a current-current interaction thought of as only mediated by *virtual* photons (the symmetric part of the propagator, $\bar{D}$) -- contains the possibility of a *real* photon (the positive energy part of the propagator, $D_1$). The latter is there in the presence of an absorber response because it represents the possibility of a transaction, as described above; and *the 'real photon' is just the actualized transaction*. In standard quantum field theory, the Feynman propagator is a hybrid form that has to include both interval (virtual) propagation and the possibility of real propagation, because it has no way of distinguishing between the two at the level of basic field propagation. In contrast, the direct-action picture can distinguish these two situations by describing genuinely internal (virtual) propagation by the time-symmetric propagator $\bar{D}$, while the real photon, which can be described as a 'free field' $D_1$ from the standpoint of quantum field theory, arises only in the context of an absorber response. This account is not clearly stated in Davies' presentations because his aim is to show the equivalence between the two theories, and his focus is on showing that including responses from absorbers provides for this equivalence. So in his analysis, he implicitly assumes that there is always absorber response, even though this is not the case for virtual photons. (However he does note that genuinely internal photons are in fact described only by the time-symmetric propagator in his direct action picture.)

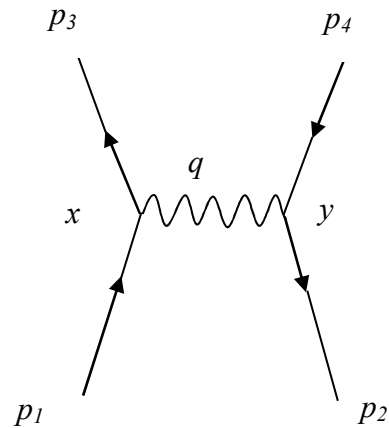

Figure 1. A virtual photon tied to an electron and positron current.

In scattering experiments such as the Bhabha scattering depicted in Figure 1, virtual photons give rise to forces between currents ('elastic collisions' of the charges). In general, such virtual photons do not satisfy energy conservation. As discussed above, according to the transactional picture, there is a small chance, proportional to $\alpha \sim (1/137)$, that *if* one of the virtual photons satisfied energy/momentum conservation for the process, it could be elevated to a 'real' photon via an actualized transaction. This would result in inelastic scattering, i.e., real energy would be deducted from one charge and added to the other. The actual probability of such a process is much smaller than $\alpha$ since it is also contingent on the energy of the initially virtual photon satisfying the conservation laws. However, it is a real predicted effect that could, in principle, be detected in the laboratory.

## 5. Energy level shifts

The interpretation of virtual photons as entities subject to promotion into real photons by way of transactions provides a natural way to understand certain computations in relativistic quantum mechanics that can seem rather artificial. For example, Sakurai (1973, Section 2-8) discusses a standard perturbative method of calculating the energy shift of a level in the hydrogen atom. (See Figure 2.) An electron in an unstable excited state *A* undergoes self-interaction by emitting and absorbing virtual photons. The emission of a virtual photon places the electron briefly in a different intermediate state *I*. The perturbation involves calculating the matrix elements corresponding to emission and absorption of the virtual photons. This gives rise to an integral whose integrand (of the usual form exhibiting the time-dependence of a quantum state, $\sim exp(-iEt)$), is an oscillating function of the variable of integration *(t)* and whose value is therefore indefinite. In order for the integral to converge, a small positive imaginary quantity *(iε)* must be added to the energy, so that the integrand becomes of the form $\sim exp(-iEt) exp(-\varepsilon t)$. Then the integral may be evaluated, yielding an energy shift with a real and imaginary part. The real part is the value of the level shift, while the imaginary part is the decay rate of the level – which resulted from the apparently *ad hoc* addition of *(iε)*. Note that both these quantities – the level shift and the decay rate -- are observable aspects of atomic levels that can be checked in the laboratory. Yet in order to get the physically valid decay rate, one has had to put in 'by hand' an imaginary quantity *(iε)* whose only apparent justification was to make an integral whose value was undefined more 'well-behaved.' Why should this *ad hoc* mathematical maneuver have any observable, verifiable physical content?

The puzzle is answered by the transactional picture: the integral may be evaluated not because of an *ad hoc* mathematical maneuver but because there are, in fact, real photons involved in the process whose existence *requires* the quantity *(iε)*. As discussed in Section 4, there is a probability of $\alpha$ that a virtual photon of energy $h\nu$, satisfying energy conservation, will be elevated to the status of a real photon. This can only come about via three distinct processes: (i) a photon 'offer wave' $|h\nu\rangle$ is emitted at one vertex (with amplitude $\sqrt{\alpha}$), and absorbed at the other vertex; (ii) an advanced confirmation wave $\langle h\nu|$ is generated at the 'absorbing' vertex,

(also with amplitude $\sqrt{\alpha}$ ); iii) the incipient transaction, with weight $\alpha$, is actualized (with probability $\alpha$). Steps (ii) and (iii) can only occur for photons satisfying energy and momentum conservation.

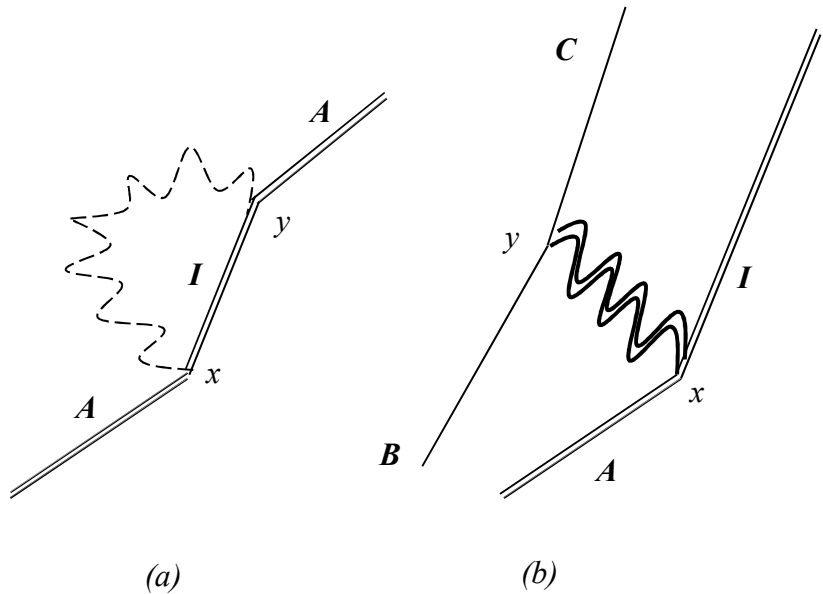

Figure 2.  (a) Self—action of an electron in atomic level $A$;
(b) A virtual photon is elevated into a real photon through the decay process.

Let us consider this process in more detail with reference to Figure 2. In this case, we are considering a hydrogen atom with a single electron in an arbitrary excited state; the electron in the potential of the nucleus is represented by a double line. Figure 2(a) depicts how the electron, initially in state $A$, is continually emitting and re-absorbing virtual photons. These are not actualized transactions, but only the possible emission and re-absorption of offer waves. Each vertex carries a factor of $\sqrt{\alpha}$ and there is a time-symmetric photon propagator $-i\overline{D}(x-y)g_{\mu\upsilon}$ connecting them.[6] The propagator describes process (i) above. Since according to the Davies theory the process is time-symmetric, there is no fact of the matter about whether $x$ or $y$ is 'earlier.' (In fact, this makes the approach simpler by not attempting to impose temporal order at the virtual propagation level.) This process contributes to the relativistic shift of the energy level, which is obtained by summing over such virtual photon contributions for all intermediate levels corresponding to arbitrary photon energies.

If the intermediate state $I$ is an allowable lower-energy state for the electron, a process such as depicted in Figure 2(b) becomes possible. Recall (see equation (3)) that in the Davies theory, the field $A_\mu(x)$ is replaced by a direct interaction between currents $(D_F(x-y) J(y))$[7]. In the transactional picture, a photon offer wave whose 4-momentum satisfies the conservations laws generates a confirmation ("response of the absorber") from at least one electron current external to the atom, depicted in Figure 2(b) as starting out in state $B$. This corresponds to process (ii) above, and (in terms of the amplitude of the process) contributes a factor of the complex conjugate of the propagator (i.e., "negative frequency part") times $\sqrt{\alpha}$. Finally, process (iii) may occur with a probability corresponding to both the above factors, where now $x^0 < y^0$ since the energy is transferred by way of the retarded, positive 'free field' $D^+(y-x)$ (cf. Davies 1971, eqn (6), where $x^0 < y^0$)[8]. This corresponds to the standard decay rate $\sim \alpha |<0|A_\nu(y)A_\mu(x)|0>|^2$ for an unstable atomic level when the photon propagator is part of the perturbation operating on the initial electron state $A$ and resulting in the final state $I$ (i.e. with zero free photons in the initial and final states).

---

[6] According to the Davies theory, these are time-symmetric, nonquantized fields, so the VEV expression of standard QED, $<0|A_\nu(y)A_\mu(x)|0>$, is replaced by the explicit time-symmetric propagator. The use of Feynman diagrams in connection with the Davies theory is legitimate: Davies (1972) applies his formulation explicitly to Feynman diagrams, and discusses in particular the lowest order self-action Feynman diagram in terms of emission and absorption by the same current (although that 'current' in his Figure 4 represents *all* the currents in a light-tight box). He analyzes the emission of a photon by one current A and its absorption by a different current B by summing over a complete set of 'fictitious photons'. This technique is necessary in his picture since he does not include the transactional process, which itself establishes independence of the emission and absorption. In the transactional picture one simply notes that the absorption is accompanied by an absorber response, and it is the latter that establishes temporal order. The actualization of a transaction is inherently unpredictable, but where such a process takes place, there is no need to sum over a set of fictitious photons to establish the independence of the emission and the absorption.

[7] Recall that the Feynman propagator, as opposed to the time-symmetric propagator, is only explicitly necessary if there are 'real' photons present. In our approach these are just the Fock states leading to actualized transactions.

[8] Note that the 'free field' part results from the sum of the retarded and advanced fields from *all* responding absorbers, even though only one actually receives the real energy conveyed by the free field. Cf. Davies (1971, eqn (28)).

It is worth noting that despite the fact that both the expression for the energy level shift $\Delta E_A$ of a level $A$ and for the decay rate $\Gamma_A$ of level $A$ to level $I$ involve the absolute square of a transition amplitude (generically $<A|H'_{AI}|I>$, where $H'_{AI}$ is the perturbing Hamiltonian between the two states), this quantity has a different physical meaning in each of the two applications. As noted above, for virtual photons as depicted in Fig. 2(a), there are *virtual* transitions of the electron from state $A$ to state $I$ and back again ($A \rightarrow I \rightarrow A$) based on the unitary action of the perturbing Hamiltonian on the electron state (offer wave), but no confirmations that could give rise to a transaction and resulting (nonunitary) real photon (i.e., real transfer of energy). The quantities representing the first process are therefore just products of successive amplitudes, divided by the difference between the energy difference of the two states and the energy of the virtual photon, so that the overall transition amplitude $A \rightarrow I \rightarrow A$ is diminished in proportion to the degree to which energy conservation is violated. In contrast, the process depicted in Fig. 2(b) involves confirmation of an offer wave and resulting incipient transaction, a non-unitary process: $A \Leftrightarrow I$.

Consider once more the presentation in Sakurai (1973, Section 2.8) in which a small imaginary quantity $i\varepsilon$ is added to the energy $E$ in the phase of the integrand, formally as follows:

$$exp(i(E + i\varepsilon) t = \quad exp(iEt) \, exp(-\varepsilon \, t) \tag{5}$$

so that the integrand acquires a purely real factor $exp(-\varepsilon t)$; and that this *ad hoc* quantity then turns out to be the empirically detectable decay rate $\Gamma$ for the level in question. Recall also that the sum over the intermediate states $I$ includes photons of *all* energies, including ones that satisfy energy conservation. But then one must include an additional term – " the imaginary part of the energy shift" – for only those photons that satisfy energy conservation for the difference between allowed levels. In the standard account, not only is the imaginary component put in 'by hand' but it also appears to count some of the photons twice (since photons that happen to satisfy energy conservation are included in the double sum over all photons and all intermediate states *I*). Moreover, the second counting has a distinct mathematical character (imaginary as opposed to real, at least in terms of the quantity $i\varepsilon$). In the transactional picture, all of this gains a clear physical basis: the real and imaginary parts of the energy shift correspond to fundamentally different physical processes. One must 'double count' photons because there is no way to predict which virtual photons, of the eligible ones, will be elevated to 'real' photons and thus become part of the decay rate (imaginary part of the energy shift). Note further that, in view of (5) above, that the 'imaginary' part of the energy shift actually corresponds to a real quantity – and that real quantities are the ones that are empirically observable.

We see, therefore, that the ability of an atomic excited state to decay may be naturally understood, without resort to *ad hoc* computational maneuvers, via the transactional picture. Real photons are simply virtual photons that have been elevated to the level of 'reality' by satisfying energy and momentum conservation and surviving the odds ($\sim \alpha \cong 1/137$) so that they may function to transfer energy from one place to another and thus result in empirical detection.

One final remark is in order concerning Davies' theoretical formulation. In the third of his three papers presenting his quantum extension of the W12F theory, he concludes with the following comment:

"This then represents the quantum generalization of the classical Wheeler12Feynman absorber theory, although it is puzzling that *S* is not unitary for all fermion states in this case (it is, of course, unitary for the light tight box)." (Davies 1972, p. 1035).

Unitarity (conservation of probability in this context) is preserved with the understanding that no real photon will exist in the first place unless there is absorber response--that is a condition for reality of the photon. In effect, the 'offer wave' (i.e., the Fock State heralding a real photon) and 'confirmation wave' co-create each other; there is never one without the other. Thus, the 'light tight box' condition in effect always applies, and does not have to be imposed as an *ad hoc* boundary condition.

7. Conclusion.

It has been pointed out that the imaginary part of the energy level shift obtained through an ad hoc procedure in standard relativistic quantum mechanics has a natural interpretation in a process predicted by the relativistic extension of Cramer's Transactional Interpretation, the Possibilist Transactional Interpretation (PTI) (Kastner 2012a,b). This process provides for a natural understanding of the distinction between real and virtual photons, in which virtual photons are unconfirmed field connections (the time-symmetric propagator), and real photons are Fock states ('offer waves') resulting only when there is absorber response.